\documentclass[twocolumn,english,showpacs,pra]{revtex4}
\usepackage[T1]{fontenc}
\usepackage[latin9]{inputenc}
\usepackage{float}
\usepackage{amsmath}
\usepackage{graphicx}
\usepackage{amssymb}
\usepackage{esint}

\makeatletter
\@ifundefined{textcolor}{}
{%
 \definecolor{BLACK}{gray}{0}
 \definecolor{WHITE}{gray}{1}
 \definecolor{RED}{rgb}{1,0,0}
 \definecolor{GREEN}{rgb}{0,1,0}
 \definecolor{BLUE}{rgb}{0,0,1}
 \definecolor{CYAN}{cmyk}{1,0,0,0}
 \definecolor{MAGENTA}{cmyk}{0,1,0,0}
 \definecolor{YELLOW}{cmyk}{0,0,1,0}
 }

\usepackage{float}

\makeatother

\usepackage{babel}

\begin{document}

\title{Developing a robust approach to implementing non-Abelian anyons and topological quantum computing in a modified Kitaev honeycomb lattice model}

\author{Haitan Xu}

\author{J. M. Taylor}

\affiliation{Joint Quantum Institute, University of Maryland, College Park, MD
20742, and National Institute of Standards and Technology, Gaithersburg,
MD 20899}
\begin{abstract}
Quantum computation provides a unique opportunity to explore new regimes of physical systems through the creation of non-trivial quantum states far outside of the classical limit.  However, such computation is remarkably sensitive to noise and undergoes rapid dephasing in most cases.  One potential solution to these prosaic concerns is to encode and process the information using topological manipulations of so-called anyons, particles in two dimensions with non-Abelian statistics.  Unfortunately, practical implementation of such a topological system remains far from complete, both in terms of physical methods but also in terms of connecting the underlying topological field theory with a specific physical model, including the imperfections expected in any realistic device.  Here we develop a complete picture of such topological quantum computation using a variation of the  Kitaev honeycomb Hamiltonian as the basis for our approach.  We show the robustness of this system against noise, confirm the non-Abelian statistics of the quasi-particles to be Ising anyons, and develop new techniques for turning topological information into measurable spin quantities.
\end{abstract}
%JMT: need to add PACS
\maketitle

Topological quantum computation shows promise for providing an approach to implementing logical quantum information operations that are robust against local noise sources due to a combination of topology, temperature, and a gapped Hamiltonian with appropriate below-gap statistics~\cite{freedman02,nayakRMP,bravyi.2010.093512,kitaev03}.  However, open questions remain for any physical implementation of such a topological system, as robustness is only guaranteed in the thermodynamic limit, and in many cases methods for implementing key operations such as preparation and measurement of quasi-particles remains at best uncertain~\cite{nayakRMP,kitaev03,kells.2008.240404,zhang.2007.18415---18420,PhysRevLett.91.090402,sarma:166802}.  Furthermore, there does not currently exist an approach for connecting potential real-world errors introduced by defects in Hamiltonian implementation and in operations to errors within the topological sector, though preliminary efforts to examine errors have commenced~\cite{chesi.2010.022305,dhochak.2010.117201}.

In this work we develop a specific physical picture for computation and noise using a variant of the celebrated Kitaev honeycomb model~\cite{Kitaev06}.  Our work follows in the spirit of recent attempts to explore explicit architectures for topological quantum computation~\cite{blueprint}, but here we focus instead upon finding appropriate methods for mapping known solutions from circuit-based fault-tolerant quantum computing to the case of finite-size, finite-time topological computing.  In one crucial respect, our work departs from prior analyses of the Kitaev honeycomb which have focused directly upon the extended version of the model, with explicit 3 spin interactions~\cite{shi.2009.134431,Yu2008,kells.2008.240404}, or upon small size (16 spin) versions~\cite{chen.2010.235131}.  Instead, we choose to introduce a gap in the so-called gapless phase using two-spin interactions with ancillary, gap-protected spin ``gadgets''~\cite{jordan08}, rather than the original suggestion~\cite{Kitaev06} of a homogeneous external magnetic field.  With this addition, the gap introduced is the first perturbative correction regardless of the number of vortices (topological excitations) existing in the system, in contrast to the case of a magnetic field-induced gap.  We remark that our approach could be implemented experimentally in manner similar to Ref.~\cite{PhysRevLett.91.090402}.

We proceed by first developing our approach for implementing the underlying, extended Kitaev model comprising two-spin interactions only and include all effects in the relevant perturbation theory up to the order of the induced gap.  We then detail how adiabatic control of the parameters of the Hamiltonian enables the creation and braiding of anyons~\cite{wilczek.1982.1144} sufficient for observing Ising anyon statistics using numerical approximations based upon the exact solution provided first by Kitaev and in the spirit of Ref.~\cite{lahtinen.2009.093027}.  We describe an explicit method for taking advantage of non-topological sectors, such as when vortices are adjacent to each other, to convert topological information into spin observables.  Finally, we analyze how control errors can lead to a quantum random walk, and show this is suppressed exponentially as vortex density decreases.  Thus we provide a complete picture, including errors, for finite-size, finite-time topological quantum computing.

\begin{figure}
\includegraphics[width=6cm]{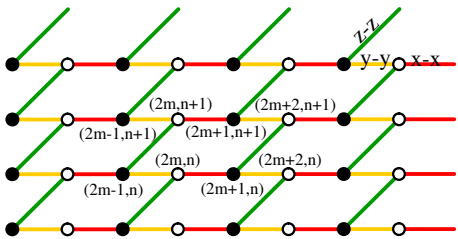}

\caption{\label{Flo:lattice} Kitaev honeycomb lattice shown as a tilted brickwall, with each brick corresponding to a single honeycomb.  Each black and white circle corresponds to a spin, with a black-white pair forming the unit cell.  Interactions between spins are determined by the color of the link (red, $x-x$; yellow, $y-y$; green, $z-z$).}

\end{figure}

We start with the Kitaev honeycomb model~\cite{Kitaev06}, which comprises a honeycomb lattice of spin-1/2 particles with two-spin interactions along the links of the honeycomb (Fig.~\ref{Flo:lattice}).  For simplicity of labeling, we deform the honeycomb into an isometric brickwall structure, such that each spin is labeled on a cartesian grid with two spins of the unit cell adjacent on the $x$ axis.  Thus $N \times N$ unit cells map to $N$ lines of $2N$ spins with $z-z$ couplings between the lines and alternating $x-x$ and $y-y$ couplings along the lines.  The corresponding Hamiltonian is
\begin{eqnarray}
H_{\text{0}}
%& = & -J_{x}\sum_{x-links}\sigma_{j}^{x}\sigma_{k}^{x}-J_{y}\sum_{y-links}\sigma_{j}^{y}\sigma_{k}^{y}\nonumber \\
 %&  & -J_{z}\sum_{z-links}\sigma_{j}^{z}\sigma_{k}^{z}\nonumber \\
 & = & -\sum_{m=1}^{M}\sum_{n=1}^{N}(J_{x}\sigma_{2m,n}^{x}\sigma_{2m+1,n}^{x}\nonumber \\
 &  & +J_{y}\sigma_{2m-1,n}^{y}\sigma_{2m,n}^{y}\nonumber \\
 &  & +J_{z}\sigma_{2m-1,n}^{z}\sigma_{2m,n+1}^{z})\label{eq:honeycomb}
 \end{eqnarray}
One key advantage of this model is the ability to exactly diagonalize the Hamiltonian by appropriate use of constants of motion (vortex operators) and use of Majorana fermions.

The Kitaev model is known to have two different topological phases, characterized by the relative values of the coupling constants. If $J_{\alpha}+J_{\beta}>J_{\gamma}$
does not hold for all permutations of $x,y,z$, then the model supports
Abelian anyon excitations and has gapped fermionic modes. Otherwise
it has been suggested that the Kitaev model should support non-Abelian vortices but have gapless fermionic modes~\cite{Kitaev06}.  In this paper we will focus on this non-Abelian (gapless) phase.

The anyonic excitations associated with the topology can be labeled by defining the vortex operators: $W_{m,n}=\sigma_{2m-1,n}^{x}\sigma_{2m,n+1}^{y}\sigma_{2m+1,n+1}^{z}\sigma_{2m+2,n+1}^{x}\sigma_{2m+1,n}^{y}\sigma_{2m,n}^{z}$,
which has eigenvalues $\pm1$ and commute with the Hamiltonian. If the expectation value of $W_{m,n}$
is -1, then there is a vortex at the plaquette $(m,n)$. For simplicity,
one can choose $J_{x}=J_{y}=J_{z}=J$ for the gapless phase.

In the gapless phase, the vortices do not have well-defined statistics as
they are strongly coupled to the fermionic modes near the singularity
of the spectrum. To open a gap in the gapless phase, and thus enable a topologically protected non-Abelian anyonic system, one approach is to
break the time-reversal symmetry of the model~\cite{Kitaev06} by the introduction of a homogeneous magnetic field $V_h = \sum_{m,n} h (\sigma^x_{m,n}+\sigma^y_{m,n}+\sigma^z_{m,n})$.  We will choose a different approach, but first we motivate our choice.

Specifically, there are several often overlooked complications to the introduction of the magnetic field.  Kitaev considered the case where the $N^2$ vortex operators had eigenvalues of $+1$, i.e., the vortex-free sector.  In this case, the introduction of the magnetic field has no effect until third order in perturbation theory, leading to the introduction of three-spin interactions with a strength $K\sim h^3 / J^2$, which create a gap of similar size.  In contrast, when there are vortices in the system, as would occur during any braiding operation, additional terms enter at even first order in perturbation theory.
This can be seen by examination of the situation with two vortices separated at a lattice distance much larger than the correlation length $\xi \sim J^3 / h^3$.  Within this space, of fixed vortex number,
the vortices are allowed to hop. To be explicit, suppose there is
a vortex on the left side of the link $\sigma_{2m-1,n}^{z}\sigma_{2m,n+1}^{z}$
and far away from the other vortices (with the state denoted by $|\psi_{1}\rangle$),
we know that $\langle\psi_{1}|W_{m-1,n}|\psi_{1}\rangle=-1$ while
$\langle\psi_{1}|W_{m,n}|\psi_{1}\rangle=+1$. The magnetic field introduces
terms, such as $h_{z}\sigma_{2m-1,n}^{z}$,
$\frac{h_{x}h_{y}}{J}\sigma_{2m-1,n}^{x}\sigma_{2m,n}^{y}$
and $\frac{h_{x}h_{y}h_{z}}{J^{2}}\sigma_{2m-1,n}^{x}\sigma_{2m,n}^{z}\sigma_{2m+1,n}^{y}$
(resulting from first, second and third order perturbation respectively),
are of higher or of the same order as the gap terms, no matter how
small the magnetic field is.
E.g., the term $h_{z}\sigma_{2m-1,n}^{z}$ induces hoping by moving the vortex to the other side of the link $\sigma_{2m-1,n}^{z}\sigma_{2m,n+1}^{z}$.
Specifically, since $\langle\psi_{1}|\sigma_{2m-1,n}^{z}W_{m-1,n}\sigma_{2m-1,n}^{z}|\psi_{1}\rangle=+1$
while $\langle\psi_{1}|\sigma_{2m-1,n}^{z}W_{m,n}\sigma_{2m-1,n}^{z}|\psi_{1}\rangle=-1$,
$|\psi_{2}\rangle=\sigma_{2m-1,n}^{z}|\psi_{1}\rangle$ has a vortex
at the plaquette $(m,n)$ instead of $(m-1,n)$. This hopping occurs at first order in perturbation theory, while the gap is {\rm third} order.  This is particularly challenging to work with, even if the gap remains formally correct, as in order to manipulate
the vortices and to avoid unwanted braiding operations, one needs
to know and fix the positions of the vortices. The vortex dynamics induced by the homogeneous magnetic field thus
appear to go gainst the efforts for manipulating and observing non-Abelian
anyons experimentally.

\begin{figure}
\begin{centering}
\includegraphics[width=6cm]{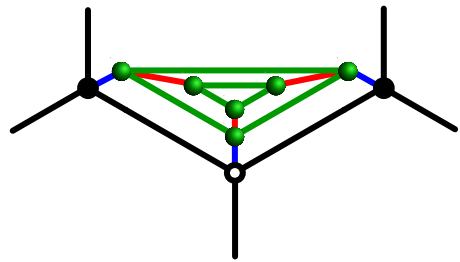}
\par\end{centering}

\caption{
Example ``gadget'' for implementing the effective three-spin interaction.  6 ancillary spins are introduced, with strong $z-z$ (green) and weak $x-x$ (red) interactions.  The ancillary system is in turn weakly coupled to the underlying honeycomb sites with with $x-x$, $x-z$, and $x-y$ interactions (blue, going clockwise).  After removal of the high-energy subspace of the gadget, the effect is to produce three-spin interactions between the honeycomb spins.
\label{fig:gadget}}

\end{figure}

Here we suggest a different approach to generate the same three-body terms
in the generalized Kitaev model %(Eq.~(\ref{eq:generalizedKitaevmodel}))
using two-body interactions only.  Our choice will lead to the necessary three-spin interactions to introduce a gap in the so-called gapless regime as the first perturbative correction that is non-zero, regardless of vortex sector.  Specifically, we wish to integrate out the auxiliary spins and recover the generalized Kitaev model:
\begin{eqnarray}
H(t) & = & \sum_{m=1}^{M}\sum_{n=1}^{N}(J_{m,n}^{x}(t)\sigma_{2m,n}^{x}\sigma_{2m+1,n}^{x}\nonumber \\
 &  & +J_{m,n}^{y}(t)\sigma_{2m-1,n}^{y}\sigma_{2m,n}^{y}\nonumber \\
 &  & +J_{m,n}^{z}(t)\sigma_{2m-1,n}^{z}\sigma_{2m,n+1}^{z}\nonumber \\
 &  & +K_{m,n}^{1}(t)\sigma_{2m+1,n+1}^{x}\sigma_{2m,n+1}^{y}\sigma_{2m-1,n}^{z}\nonumber \\
 &  & +K_{m,n}^{2}(t)\sigma_{2m,n+1}^{x}\sigma_{2m+2,n+1}^{y}\sigma_{2m+1,n+1}^{z}\nonumber \\
 &  & +K_{m,n}^{3}(t)\sigma_{2m+2,n+1}^{x}\sigma_{2m+1,n+1}^{y}\sigma_{2m+1,n}^{z}\nonumber \\
 &  & +K_{m,n}^{4}(t)\sigma_{2m,n}^{x}\sigma_{2m+1,n}^{y}\sigma_{2m+2,n+1}^{z}\nonumber \\
 &  & +K_{m,n}^{5}(t)\sigma_{2m+1,n}^{x}\sigma_{2m-1,n}^{y}\sigma_{2m,n}^{z}\nonumber \\
 &  & +K_{m,n}^{6}(t)\sigma_{2m-1,n}^{x}\sigma_{2m,n}^{y}\sigma_{2m+1,n+1}^{z})\label{eq:generalizedKitaevmodel}\end{eqnarray}

We use the idea of
perturbative gadgets~\cite{Kempe04,jordan08} to create each 3 spin interaction in Eq.~(\ref{eq:generalizedKitaevmodel}).  This leads to an overhead of 6 additional spins for every 3-spin interaction on the honeycomb.  Furthermore, we extend the gadget approach by adding an additional protection against local noise, thus maintaining the gap- and topological-protection.  Specifically, rather than using a single set of ancillary spins per interaction term, we
% the two cat
%states, $\frac{1}{\sqrt{2}}|\uparrow\uparrow\uparrow\pm\downarrow\downarrow\downarrow\rangle$,
%of the auxilirary spins, are not topologically protected and can change
%from one state to the other without raising energy. This could be
%a serious problem as $K$ may change continuously from one sign to
%the other sign, which certainly introduces errors to the model Hamiltonian.
%
%In order to make the model Hamiltonian topologically stable to small
%finite local noises,
use two sets of auxiliary spins to generate each three-body interaction (more sets resulting in better protection). The two
sets of auxiliary spins are coupled together, as shown in Fig.~\ref{fig:gadget},
\begin{equation}
\begin{split}H_{\text{aux},jkl}= & -\frac{A}{2}(\tilde{\sigma}_{j}^{z}\tilde{\sigma}_{k}^{z}+\tilde{\sigma}_{k}^{z}\tilde{\sigma}_{l}^{z}+\tilde{\sigma}_{l}^{z}\tilde{\sigma}_{j}^{z}\\
 & +\bar{\sigma}_{j}^{z}\bar{\sigma}_{k}^{z}+\bar{\sigma}_{k}^{z}\bar{\sigma}_{l}^{z}+\bar{\sigma}_{l}^{z}\bar{\sigma}_{j}^{z})\\
 & -\tilde{\lambda}(\tilde{\sigma}_{j}^{x}\bar{\sigma}_{j}^{x}+\tilde{\sigma}_{k}^{x}\bar{\sigma}_{k}^{x}+\tilde{\sigma}_{l}^{x}\bar{\sigma}_{l}^{x}).\end{split}
\label{eq:auxiliary}\end{equation}
Each set of three spins have ferromagnetic, Ising-like couplings ($z$-$z$). The two sets of spins are coupled to each other perturbatively
through $x$-$x$ bonds. The ground state manifold of $H_{\text{aux},jkl}$
(for $\tilde{\lambda}=0$) can be written in terms of four `cat' states, \begin{equation}
\frac{1}{2}|\tilde{\uparrow}\tilde{\uparrow}\tilde{\uparrow}\pm\tilde{\downarrow}\tilde{\downarrow}\tilde{\downarrow}\rangle\otimes|\bar{\uparrow}\bar{\uparrow}\bar{\uparrow}\pm\bar{\uparrow}\bar{\uparrow}\bar{\uparrow}\rangle\label{eq:-1}\end{equation}
The $x-x$ bonds breaks the degeneracy of the cat states. The
states $\frac{1}{2}|\tilde{\uparrow}\tilde{\uparrow}\tilde{\uparrow}+\tilde{\downarrow}\tilde{\downarrow}\tilde{\downarrow}\rangle\otimes|\bar{\uparrow}\bar{\uparrow}\bar{\uparrow}+\bar{\uparrow}\bar{\uparrow}\bar{\uparrow}\rangle$
and $\frac{1}{2}|\tilde{\uparrow}\tilde{\uparrow}\tilde{\uparrow}-\tilde{\downarrow}\tilde{\downarrow}\tilde{\downarrow}\rangle\otimes|\bar{\uparrow}\bar{\uparrow}\bar{\uparrow}-\bar{\uparrow}\bar{\uparrow}\bar{\uparrow}\rangle$
are still degenerate ground states of the effective Hamiltonian, but
separated by an energy $\sim3\frac{\tilde{\lambda}^{3}}{A^{2}}$ from
the other cat states. The two ground states cannot be continuously
changed into each other under local noise without raising the energy, and the transition between the two states is suppressed.  Furthermore, if such noise does cause a change between the two degenerate ground states, the effect on the extended Kitaev hamiltonian occurs as local noise, which should be protected against if the system exhibits topological order~\cite{bravyi.2010.093512}.
%This could be stable under noises
%smaller than the gap energy. (The auxiliary spins themselves may be
%used to do quantum computation, at least serving as stable qubits.)

Now we can use gadgets in one of the two degenerate ground states to generate three-body terms.  We couple the original spins to the auxiliary spins by \begin{equation}
H_{\text{coupl},jkl}=-\lambda(\tilde{\sigma}_{j}^{x}\sigma_{j}^{x}+\tilde{\sigma}_{k}^{x}\sigma_{k}^{y}+\tilde{\sigma}_{l}^{x}\sigma_{l}^{z}),\label{eq:-1}\end{equation}
Suppose each two sets of the auxiliary spins have already been initialized
to the state $\frac{1}{2}|\tilde{\uparrow}\tilde{\uparrow}\tilde{\uparrow}+\tilde{\downarrow}\tilde{\downarrow}\tilde{\downarrow}\rangle\otimes|\bar{\uparrow}\bar{\uparrow}\bar{\uparrow}+\bar{\uparrow}\bar{\uparrow}\bar{\uparrow}\rangle$.
Using third order perturbation theory, we obtain \begin{equation}
H_{\text{eff}}=-J\sum_{\alpha\textrm{-}links}\sigma_{j}^{\alpha}\sigma_{k}^{\alpha}-\frac{3}{2}\frac{\lambda^{3}}{A^{2}}\sum_{\{j,k,l\}\in TI}\sigma_{j}^{x}\sigma_{k}^{y}\sigma_{l}^{z}.\label{eq:honeycomb2}\end{equation}
Compared with Eq.~(\ref{eq:generalizedKitaevmodel}), we have $K=\frac{3}{2}\frac{\lambda^{3}}{A^{2}}$.
To open a large spectral gap comparable to $J$ in the gapless phase
while preserving the stability of the cat states of auxiliary spins,
we require $J\sim\frac{\lambda^{3}}{A^{2}}$ and $\tilde{\lambda}^{3}\gg\lambda^{3}$
(rather than $\tilde{\lambda}\gg\lambda$). The interaction between
vortices can be neglected as long as they are separated from each
other by a distance $\gg|\frac{J}{K}|\sim1$.  Finally, control of the three-spin interaction is enabled by changing a single interaction between the gadget spins and the honeycomb spins.
%We remark that this distance could be subwhich is much shorter
%than the case in Kitaev's paper using magnetic field).

%

%Disorder effect in Kitaev model?

%In the following, we study the stability of the revised Kitaev model against local errors. This problem is quite difficult to solve even by numerical calculation. Nevertheless, we can reduce the difficulty of the problem by studying local errors which change the coupling constants to learn the stability of Kitaev model. The following discussion is based on numerical calculation of a honeycomb lattice of size $18\times18$.

In order to study the topological properties of the Kitaev honeycomb
model, we generalize it by allowing for the coupling terms $J_{m,n}^{\mu}, K_{m,n}^k$ externally controlled, and variable, parameters.
%, i.e., we study the following generalized
%Kitaev model\begin{eqnarray}
%H(t) & = & \sum_{m=1}^{M}\sum_{n=1}^{N}(J_{m,n}^{x}(t)\sigma_{2m,n}^{x}\sigma_{2m+1,n}^{x}\nonumber \\
% &  & +J_{m,n}^{y}(t)\sigma_{2m-1,n}^{y}\sigma_{2m,n}^{y}\nonumber \\
% &  & +J_{m,n}^{z}(t)\sigma_{2m-1,n}^{z}\sigma_{2m,n+1}^{z}\nonumber \\
% &  & +K_{m,n}^{1}(t)\sigma_{2m+1,n+1}^{x}\sigma_{2m,n+1}^{y}\sigma_{2m-1,n}^{z}\nonumber \\
% &  & +K_{m,n}^{2}(t)\sigma_{2m,n+1}^{x}\sigma_{2m+2,n+1}^{y}\sigma_{2m+1,n+1}^{z}\nonumber \\
% &  & +K_{m,n}^{3}(t)\sigma_{2m+2,n+1}^{x}\sigma_{2m+1,n+1}^{y}\sigma_{2m+1,n}^{z}\nonumber \\
% &  & +K_{m,n}^{4}(t)\sigma_{2m,n}^{x}\sigma_{2m+1,n}^{y}\sigma_{2m+2,n+1}^{z}\nonumber \\
% &  & +K_{m,n}^{5}(t)\sigma_{2m+1,n}^{x}\sigma_{2m-1,n}^{y}\sigma_{2m,n}^{z}\nonumber \\
% &  & +K_{m,n}^{6}(t)\sigma_{2m-1,n}^{x}\sigma_{2m,n}^{y}\sigma_{2m+1,n+1}^{z})\label{eq:generalizedKitaevmodel}\end{eqnarray}
%defined on a honeycomb lattice. We will discuss how to generate the
%controllable three-body interactions from two-body interactions later
%in this paper.
The generalized Kitaev model can be simplified by spin-fermion
transformation ($\sigma_i^\alpha=i b_i^\alpha c_i$, where $b_i^\alpha$ and $c_i$ are Majorana fermions)~\cite{Kitaev06}, which results in the Hamiltonian
\begin{equation}
H(t)=\frac{\mathrm{i}}{4}\sum_{j,k}A_{jk}(t)c_{j}c_{k},\label{eq:}\end{equation}
where $A_{j,k}(t)=-2J_{p_{jk}}^{\alpha_{jk}}(t)u_{jk}+2K_{p_{jk}}^{\beta_{jk}}(t)u_{jl}u_{lk}$
if $j,k$ are linked directly or through a third spin $l$, otherwise
0. We can choose the initial coupling parameters $J_{p_{jk}}^{\alpha_{jk}}(0)=J_{0}$,
and $K_{p_{jk}}^{\beta_{jk}}(0)=K_{0}$. The link operator $u_{jk}=ib_{j}^{\alpha_{jk}}b_{k}^{\alpha_{jk}}$
has eigenvalues $\pm1$. The vortex operator is $W_{p}=\prod u_{jk}$.

Suppose the system is initialized to the vortex-free ground state.
We can adiabatically change some of the coupling parameters $J_{p_{jk}}^{\alpha_{jk}}(t)$
and $K_{p_{jk}}^{\beta_{jk}}(t)$.  This does not change the
actual vortex configuration, as vortex operators $W$ do not change their expectation value.  However, it does change the fermion spectrum, and the best description is given by considering the same model without vortices but with a change of sign of appropriate $J$'s and $K$'s as suggested in earlier work~\cite{lahtinen.2009.093027}.  These equivalent descriptions we label pseudo-vortices for this work, where the key distinction is that measurement of the vortex operator does not reveal pseudo-vortices.  We remark
here that the equivalence between changing $J$ and $u$ was pointed
out in Kitaev's paper to show that the ground state energy does not
depend on the signs of $J$~\cite{Kitaev06}.
%As we are writing up
%this paper, we notice that V. Lahtinen and J. K. Pachos also used
%this equivalence to study the non-Abelian statistics of vortices in
%Kitaev model~\cite{lahtinen.2009.093027}, but the details in this paper are
%quite different from their paper.

\begin{figure}
\centering{}\includegraphics[width=8.5cm]{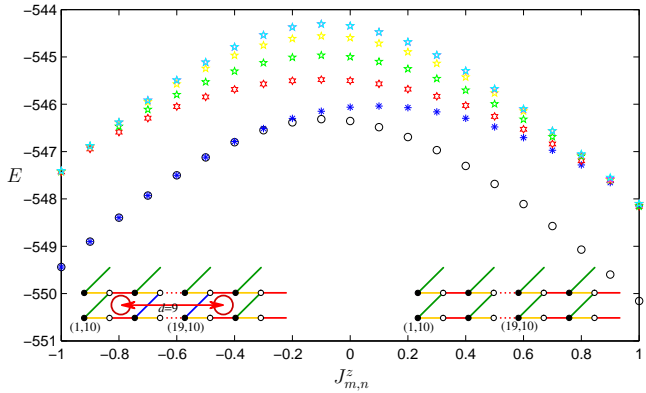}
\caption{\label{fig:Spectrum1}Spectrum of lowest energy levels vs the value
of $J_{m,n}^{z}$ for creating a pair of vortices on the top right of the spins at $(1,10)$ and $(19,10)$ in an $18\times18$ lattice (two spins per plaquette), with periodic boundary
condition, $J_{0}=1$ and $K_{0}=0.2$.}
\end{figure}

To create a pair of pseudo-vortices at plaquettes $(m_{1},n)$ and $(m_{2},n)$
in the horizontal direction, we can continuously change the sign of all of
the $J_{m,n}^{z}$, $K_{m,n}^{1,6}$ ($m_{1}+1<m<m_{2}$), and $K_{m,n}^{3,4}$
($m_{1}<m<m_{2}-1$) between the two pseudo-vortices. We show in Fig.~\ref{fig:Spectrum1}
the spectrum of lowest energy levels versus the value of $J_{m,n}^{z}$
for an $18\times18$ lattice (two spins per plaquette, choosing $m_{1}=1$, $m_{2}=10$, and
$n=10$), with periodic boundary condition, $J_{0}=1$ and $K_{0}=0.2$.
There is a (almost) zero energy fermion mode when we have changed
the sign of the relevant parameters between the two vortices.  When
the size of the system and the distance between vortices goes to infinity,
the first fermion mode should be of exactly zero energy--in fact, this first fermion mode corresponds to the topological degeneracy of the two-vortex sector.  It is also
interesting to observe how a nonzero fermion mode becomes a zero mode.
In fact, all the higher modes come down to replace their lower neighbor
modes. %We can similarly create more vortices.

\begin{figure}
\centering{}\includegraphics[width=8.5cm]{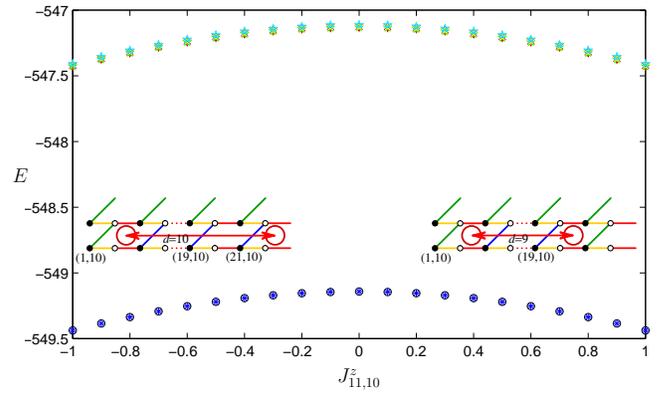}
\caption{\label{fig:spectrum2}Spectrum of lowest energy levels vs the value
of $J_{11,10}^{z}$ for moving a vortex from the top right of the spin at $(19,10)$ to the top right of the spin at $(21,11)$
in an $18\times18$ lattice. }

\end{figure}

\begin{figure}

\centering{}\includegraphics[width=8.5cm]{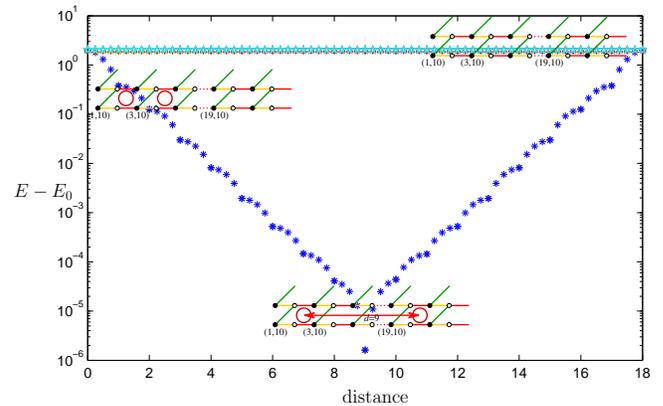}
\caption{\label{fig:spectrum3} Excited state spectrum vs distance between two vortices in
the same horizontal line by gradually changing coupling parameters to move one vortex of the pair.
Here we have set the ground state energy to 0. The horizontal axis
is the distance between two vortices. The points between integer distances correspond to the intermediate states when moving vortex from site to site. When the distance is 0 (or 18),
there are actually no vortices due to periodic boundary conditions.  Note the exponential suppression of residual energy difference as a function of distance, consistent with predictions for the topological sector.}

\end{figure}

To move a pseudo-vortex, we can change the coupling parameters of the relevant
links between the initial position and the final position. For example,
to move one pseudo-vortex at the plaquette ($m,n$) in the horizontal direction by one
site to ($m+1,n$), we continuously change the sign of the coupling
parameters $J_{m+1,n}^{z}$, $K_{m+1,n}^{1,6}$, and $K_{m,n}^{3,4}$.
We show in Fig.~\ref{fig:spectrum2} the spectrum of lowest energy
levels vs the value of $J_{11,10}^{z}$ for moving a pseudo-vortex from the plaquette $(10,10)$
to $(10,11)$ in an $18\times18$ lattice. As expected, the spectrum
remains almost the same after moving the pseudo-vortex. Again, when the
size of the system and the distance between vortices goes to infinity,
the spectrum will remain exactly the same. We also give the spectrum
versus distance between two vortices in the same horizontal line by gradually
changing sequentially the $J$ and $K$ coupling parameters for each lattice site, as shown in Fig.~\ref{fig:spectrum3}.
The gap between the zero fermionic mode and the ground state decreases
exponentially with increasing distance between the two vortices, confirming the supposition that decreasing the density of vortices leads to an exponential improvement in their degeneracy.

\begin{figure}
\includegraphics[clip,width=8.5cm]{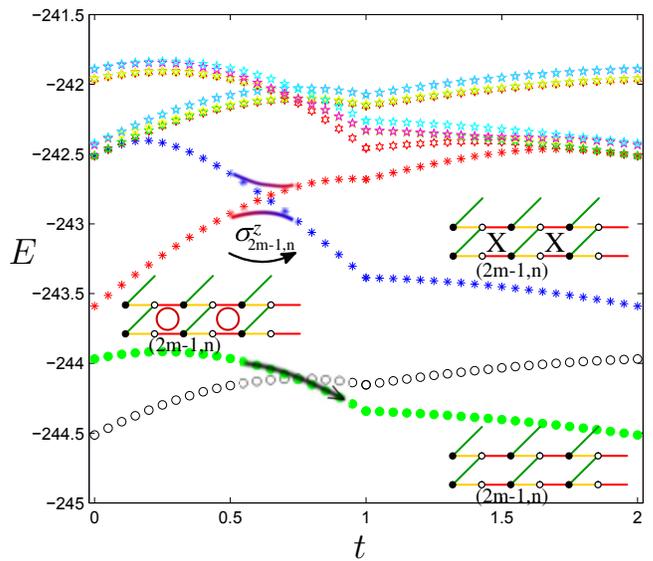}

\caption{\label{fig:fusionspectrum}
Spectrum for fusion.  From $t=0$ to $1$, the $J$ between two adjacents vortices is changed from negative to positive; from $t=1$ to $t=2$, the associated three-spin interactions' $K$s are tuned from negative to positive.  Only adiabatic evolution is considered here, corresponding to units of time much longer than $1/J$.  Offsetting these two operations leads to crossing between the two sectors (with and without real vortices) at different points in time.  Addition of a small magnetic field during one of the crossings leads to an avoided crossing and an efficient means to convert topological information into spin information which can be measured through the vortex operator $W$.}

\end{figure}

To measure the topological state of the system, we propose a method
to represent topological information by \textit{real} \textit{vortices}.
Real vortices correspond to negative eigenvalues $-1$ of vortex operator
$W_{p}=\prod u_{jk}$. In the earlier discussion about the creation and manipulation of the vortices, there
have only been pseudo-vortices in the system: those equivalent
vortices realized by changing certain coupling parameters of $J_{p_{jk}}^{\alpha_{jk}}(t)$
and $K_{p_{jk}}^{\beta_{jk}}(t)$.  Suppose we want to measure the
total topological charge of two pseudo-vortices, i.e., to measure whether
the fermionic zero mode shared by the two pseudo-vortices is occupied or not.
First we move one pseudo-vortex to be the nearest neighbor of the other pseudo-vortex.
Suppose the two pseudo-vortices are at the plaquettes $(m-1,n)$ and
$(m,n)$. At this point, the low-energy fermion modes are not degenerate any more--there
is a finite splitting between the ground state and the first excited
state for the two neighboring vortices. Next, rather than changing $J$ and $K$ together, we first adiabatically change
$J_{m,n}^{z}(t)$ from $-1$ to $1$, with the spectrum shown in Fig.~\ref{fig:fusionspectrum},
where we also give the spectrum for the vortex configuration with
two pseudo-voritices and two real vortices at the same positions. We can
see that there the degenerate point for the two ground levels of
different vortex configuration occurs at a different value of $J$ than the degenerate point for the
two first excited levels.

If, near the degenerate point for the two first excited levels, we
adiabatically turn on a local term $h_{z}\sigma_{2m-1,n}$ by, e.g.,
turning on a local magnetic field in the $z$ direction which acts
on the spin at the position $(2m-1,n)$, then the two first excited
states for the different vortex configurations will couple to each
other by $h_{z}\sigma_{2m-1,n}$. Denote the first excited state without
real vortices by $|\psi_{1}\rangle$. As $\langle\psi_{1}|\sigma_{2m-1,n}^{z}W_{m-1,n}\sigma_{2m-1,n}^{z}|\psi_{1}\rangle=-1$
and $\langle\psi_{1}|\sigma_{2m-1,n}^{z}W_{m,n}\sigma_{2m-1,n}^{z}|\psi_{1}\rangle=-1$,
$\sigma_{2m-1,n}^{z}|\psi_{1}\rangle$ has a pair of real vortices
at the plaquettes $(m-1,n)$ and $(m,n)$. The local filed results
in a finite gap proportional to the magnetic field, which we have
also calculated numerically (with a lower bound $0.46h_{z}$ corresponding
to a $12\times12$ lattice). Note that the gap does not depend on
the system size if the size is larger than the correlation length,
since the vortices are next to each other and the magnetic field is
local. When the system goes through the degenerate point for the two
first excited levels, the first excited state for the configuration
without real vortices becomes the state for the configuration with
two real vortices. Then we adiabatically turn off the magnetic field.
The ground states are not affected as the first excited states.

To finish converting pseudo-vortices to real vortices, we finally change $K_{m,n}^{1,6}$, and $K_{m-1,n}^{3,4}$ to complete the fusion of the two pseudo-vortices.
%It is interesting  to note that these two degenerate points appears at different values
%of the parameter $J_{m,n}^{z}(t)$, which is not the case if we change
%$J_{m,n}^{z}(t)$ and $K_{m,n}^{1,6}$, and $K_{m-1,n}^{3,4}$ simultaneously.
In
this way, after the fusion of the two pseudo-vortices, we obtain two
real vortices if the total topological charge of the pseudo-vortices
is 1, and obtains no real vortices if the total topological charge
of the pseudo-vortices is 0.
%We point out that there are also other
%ways to measure the topological states, e.g., we can utilize the energy
%variation of the system before and after fusion to decide whether
%the fermion mode is occupied or not. In other words, the energy to
%fuse two vortices is different for different states. The details will
%be discussed elsewhere.

%

We now conclude with a discussion of real-world constraints on topological computing which can finally be analyzed using this explicit model for braiding and measurement.
First, a honeycomb lattice with periodic boundary
condition may be difficult to realize experimentally. To obtain free boundary condition, we can simply set the links connecting the boundaries to 0.

In order to study the statistical properties of the vortices, we employ
adiabatic theory to calculate the evolution of the system as we braid
the vortices, similar in spirit to Ref.~\cite{lahtinen.2009.093027}. When a system has an almost degenerate ground subspace,
separated from the rest space by a finite gap, and if the Hamiltonian
changes adiabatically compared to the energy gap, the system will
evolve within the ground subspace throughout the process if it is
originally in the ground subspace. Let us further make the approximation
that the ground subspace is degenerate. The state of the system can
be denoted by \begin{equation}
\Psi(t)=\sum_{n}c_{n}(t)\psi_{n}(t)\exp^{\frac{-\mathrm{i}}{\hbar}\int_{0}^{t}\epsilon(t')dt},\label{eq:}\end{equation}
where $\psi_{n}(t)$ is the $n$th eigenfunction of $H(t)$ in the
ground subspace. $\exp^{\frac{-\mathrm{i}}{\hbar}\int_{0}^{t}\epsilon(t')dt}$
is an overall dynamical phase factor, while $c_{n}(t)$ contains the
geometric information. $\Psi(t)$ evolves according to \begin{equation}
\frac{\partial c_{n}(t)}{\partial t}=-\sum_{m}\langle\psi_{n}(t)|\frac{\partial\psi_{m}(t)}{\partial t}\rangle c_{m}(t).\label{eq:}\end{equation}
By diagonalizing the generalized Kitaev model, one can obtain the
spectrum and fermion modes of the system. The basis states $\psi_{n}(t)$
can thus be determined by the fermion creation/annihilation operators.
For example, when there are four vortices, there are two zero fermion
modes. The ground state without any fermion excitation $\psi_{0}(t)$
satisfies \begin{equation}
b_{k}(t)\psi_{0}(t)=0,\label{eq:}\end{equation}
for all $k$. $\psi_{1}(t)=b_{1}^{\dagger}(t)\psi_{0}(t)$, $\psi_{2}(t)=b_{2}^{\dagger}(t)\psi_{0}(t)$,
and $\psi_{3}(t)=b_{1}^{\dagger}(t)b_{2}^{\dagger}(t)\psi_{0}(t)$.
According to the adiabatic theory, we can focus on the ground subspace
expanded by $\psi_{0,1,2,3}(t)$, and thus $b_{1,2}(t)$ and $b_{1,2}^{\dagger}(t)$.
To calculate $\langle\psi_{n}(t)|\frac{\partial\psi_{m}(t)}{\partial t}\rangle$,
we can expand $b_{1,2}(t+\triangle t)$ and $b_{1,2}^{\dagger}(t+\triangle t)$
by $b_{1,2}(t)$ and $b_{1,2}^{\dagger}(t)$. Now we can study the
statistical properties of the vortices in the generalized Kitaev model.
In the following we discuss the main results obtained from numerical
calculation.

The first aim is to prove the SU(2)$_{2}$ anyonic behavior of the
vortices. At a first glance of the numerical result, e.g.,\begin{eqnarray}
B_{12} & = & \left[\begin{array}{cc}
-0.7155-0.6969\mathrm{i} & -0.0197-0.0429\mathrm{i}\\
0.0197-0.0429\mathrm{i} & -0.7155+0.6969\mathrm{i}\end{array}\right]\\
B_{23} & = & \left[\begin{array}{cc}
-0.7141-0.0510\mathrm{i} & 0.3830+0.5836\mathrm{i}\\
-0.3830+0.5836\mathrm{i} & -0.7141+0.0510\mathrm{i}\end{array}\right]\end{eqnarray}
the braiding matrices $B_{12}$ (by braiding the first two vortices)
and $B_{23}$ (by braiding the middle two vortices) are not like what
we expect from SU(2)$_{2}$ anyon model. Unlike the SU(2)$_{2}$ anyon
model, instead of each two anyons sharing one fermion mode, the four
vortices together share two fermionic modes. Even non-braiding relative
motion between two vortices can induce the evolution of the state
of the system. In fact, $B_{12}$ and $B_{23}$ for the vortices differ
approximately from those for SU(2)$_{2}$ anyon model by a unitary
transformation, \begin{eqnarray}
B_{12} & \approx & U^{\dagger}\left[\begin{array}{cc}
-\textrm{e}^{i\pi/4} & 0\\
0 & -\textrm{e}^{-i\pi/4}\end{array}\right]U\\
B_{23} & \approx & U^{\dagger}\left[\begin{array}{cc}
-\frac{1}{\sqrt{2}} & \frac{\mathrm{i}}{\sqrt{2}}\\
\frac{\mathrm{i}}{\sqrt{2}} & -\frac{1}{\sqrt{2}}\end{array}\right]U\end{eqnarray}
\begin{equation}
\textrm{where}\; U=\left[\begin{array}{cc}
0.9576+0.2859\mathrm{i} & 0.0335-0.0047\mathrm{i}\\
-0.0335-0.0047\mathrm{i} & 0.9576-0.2859\mathrm{i}\end{array}\right]\end{equation}
while the algebraic structure of the braiding operations is the same
as in the SU(2)$_{2}$ anyon model.

For a finite size system, the zero fermionic mode is usually not exactly
0, which results in the split of the degeneracy of the ground states.
Thus there will be dynamical phase errors to the braiding operations.
Even if the dynamical phase errors are corrected in some way, the
braiding matrices will still be different from the exact form. This
finite size effect is usually not discussed, and may be used to carry
out universal quantum computation with higher accuracy. There is another
way to make the generalized Kitaev model support universal topological
quantum computation. We can change the coupling parameters suddenly
or apply certain sudden magnetic pulse on a single spin to transport
a vortex. This equivalently cut the braid in the (2+1)-dimensional
spacetime, and the evolution differs from the adiabatic evolution.
The sudden and adiabatic changes of the coupling parameters together
may also support universal topological quantum computation.

Another interesting result is that there seem to be two different
topological phases (conjugate to each other) for the Kitaev model.
In order to identify the different phases, we create two vortices
and calculate the braiding matrix $B_{12}$ (which is a diagonal matrix
or a phase gate). Choosing $J_{0}=1$, and varying the value of $K_{0}$,
the phase changes from $-\mathrm{i}$ to $\mathrm{i}$ when $K$ changes
across some critical value, e.g, between 0.17485 and 0.17490. This
phase transition is accompanied by an exact zero fermionic mode at
the critical point, where the creation operator $b_{1}^{\dagger}(t)$
changes to the annihilation operator $b_{1}(t)$, and thus $\psi_{0}(t)$
and $\psi_{1}(t)$ exchange their roles with each other. Such critical
points form a subset in the parameter space of the coupling parameters
separating different phases.

We can also introduce disorder into the system by changing some coupling
parameters. For example, we can set all the coupling strength of the
links around a spin to be 0, which creates a vacancy in the system.
Our numerical calculation shows that braiding a vortex around a vacancy
point is a trivial operation.

\begin{figure}
\includegraphics[width=8.0cm]{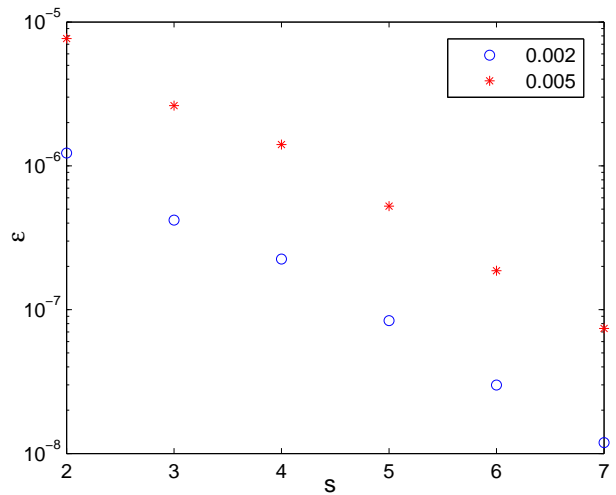}

\caption{\label{f:error} Error induced in the memory due to random variation (equally distributed within the ranges $\%0.2$ (blue circles) and $\%0.5$ (red stars)) of all coupling parameters. The Hamiltonian returns to the original Hamiltonian after 100 steps of random walks. The separation distance between the vortices is given on the bottom axes, while the error is the left axis. The error is defined to be the distance between the gates $G$ and $\tilde{G}$ with and without noise, i.e. $d\left( G,\tilde{G}\right) \equiv \Vert G - \tilde{G} \Vert = \sup _{\Vert
\psi \Vert = 1} \Vert \left( G-\tilde{G}\right) \psi \Vert$.  For the topological memory, $\tilde{G}$ is the identity operation, while $G$  is the operation with error. Increasing separation distance $s$ between the vortices leads to suppression of errors, with apparent exponential behavior, consistent with topological protection against local noise.}
%\caption{Error induced in the braid due to random variation (equally distributed within the range specified in the bottom axes) of all coupling parameters during the braid.  The maximum deviation per site is given on the bottom axes, while the error is the left axis. The error is defined to be the distance between the gates $U$ and $\tilde{U}$ with and without noise, i.e. $d\left( U,\tilde{U}\right) \equiv \Vert U - \tilde{U} \Vert = \sup _{\Vert
%\psi \Vert = 1} \Vert \left( U-\tilde{U}\right) \psi \Vert$.  Increasing minimum distance $s$ in the braid leads to suppression of errors, with apparent exponential behavior, consistent with topological protection against local noise.
%\label{f:error}}
\end{figure}

An important question about the Kitaev model is its topological stability.
It is quite difficult to discuss the effect of general noises. We
first discuss a class of noises which changes the strength of the
coupling parameters and can be calculated exactly numerically.  For example, we allow the coupling parameters to deviate randomly from the accurate values but within certain range for a system with four vortices, which form a single qubit. This results in the random walk of the topological quantum memory, though there are neither creation/annihilation nor braiding of vortices. We can see that when the system size is increased, the error of the memory is suppressed exponentially, as shown in Fig.~\ref{f:error}. We can also allow the coupling parameters to deviate randomly from
the accurate values but within certain range as we braid the vortices, which results in similar robustness behavior of the topological quantum gates.

%When the system size is fixed, the error of gates decreases approximately
%polynomially with decreasing error range of the coupling constants, as shown in Fig.~\ref{f:error}.
%For different system sizes, the error of the gates is suppressed
%with increasing system size.  Curiously, analysis of the errors shows that they are, in the basis given for our braids, primarily phase ($Z$) errors.  Specifically we find a 10-1 ratio of phase errors to bit-flip errors.  Thus, single-axes error correcting codes may be sufficient to enable tolerance against control and finite-size errors in this system~\cite{ioffe.2007.032345}.

%

Combining the above discussion with perturbation theory, we can discuss
the effects of general slow noises which act locally on the spins. There
are three types of effects. One effect preserves the number and positions
of vortices, but changes the fermion spectrum. Specifically, the two
and three orders of this effect results in the variation of link parameters,
which we have numerically studied, while higher orders introduce new
terms which are products of the lower order terms. The resulting errors
are topologically suppressed. The second effect is hopping of vortices,
which preserves the number of vortices but changes their positions.
The hopping of vortices can be suppressed by introducing six-body
interaction (as in the case of three-body interaction) which fixes
the positions of the vortices. The third effect is creation/annihilation
of vortices. The creation of vortices is suppressed by the excitation
energy, while the annihilation of vortices is suppressed by the distance
between vortices.

To further simplify the generalized Kitaev model, we can choose $K_{p_{jk}}^{2,3,5,6}(t)\equiv0$
to reduce the number of auxiliary spins. This simplification should
be compensated by larger system size.

We have suggested a complete means of analyzing the non-Abelian sector of the Kitaev honeycomb model, including the analysis of noise and the observation of topological suppression of such noise.  Our approach may also be applied to other similar systems such as in~\cite{Yao07}.  However, key open questions remain, including understanding the effect dynamical (high-frequency) noise plays.  This would correspond to terms which drive the system above the gap, and may be unavoidable in physical systems under consideration.  Nonetheless, such questions may now be answered using the framework presented herein.

\end{document}